\documentclass[11pt]{article}
\usepackage{amssymb}
\usepackage{amsmath}
\usepackage{graphicx}
\usepackage{hyperref}
\usepackage{bm}
\newcommand{\mathsym}[1]{{}}

\usepackage{graphicx}
\usepackage{rotating}
\usepackage{color}
\usepackage{cancel}

\newcommand{\bra}{\begin{array}}
\newcommand{\era}{\end{array}}
\newcommand{\beq}{\begin{equation}}
\newcommand{\eeq}{\end{equation}}
\newcommand{\beqar}{\begin{eqnarray}}
\newcommand{\eeqar}{\end{eqnarray}}

\def\BC{\bb C}
\def\_\BC{\bbi C}

\def\( {\left(}
   \def\) {\right)}
\def\[ {\left[}
\def\] {\right]}

\def\dag {{\dagger}}

\newtheorem{proposition}{Proposition}[section]

\begin{document}
\vspace{20pt}

\begin{center}

{\LARGE \bf Modified $(G'/G)$-expansion method for solving nonlinear partial differential equations\\
\medskip
 }
\vspace{15pt}

{\large Mahouton Norbert  Hounkonnou$^{*}$ and  Rolland  Finangnon Kanfon$^{\dag}$ } 

\vspace{15pt}

{\sl International Chair in Mathematical Physics and Applications\\ (ICMPA-UNESCO Chair), University of Abomey-Calavi,\\
072B.P.50, Cotonou, Rep. of Benin}\\

\vspace{5pt}
E-mails:  {$^{*}$norbert.hounkonnou@cipma.uac.bj, \sl  $^{\dag}$rolland.kanfon@cipma.uac.bj
}

\vspace{10pt}
\end{center}
\date{\today}
\begin{abstract}
This paper addresses a modified
 ($G'/G$)-expansion method 
to obtain new classes of solutions to nonlinear partial differential equations (NPDEs). The cases of Burgers, KdV and Kadomtsev-Petviashvili NPDEs 
are exhaustively  studied. Relevant graphical representations are shown in each case.
\end{abstract} 


\section{Introduction}
Over the second half of the past century, there was a flurry of activities regarding the investigation,  with a 
solid mathematical background 
and a wide range of physical 
observations and practical applications,  of nonlinearities in complex  phenomena 
governing our universe. Indeed, nonlinear equations, widely used to describe these complex phenomena,
 pervade many branches of mathematics and physics, including fluid mechanics, plasma physics, quantum
electrodynamics, solid-state physics,  nuclear and atomic physics, plasma 
waves, biology, and so on. See, for instance, \cite{A5, A6, b22} (and references therein) for more details from both theoretical and
experimental viewpoints. The knowledge of exact   solutions to these equations constitutes certainly 
one of the keys for  better understanding the nonlinearities. This explains the search for suitable analytical methods for solving nonlinear 
equations. The most popular ones are today the 
variational iteration method \cite{A11}, \cite{A13}, inverse 
scattering method \cite{A16}, Hirota bi-linear
 form\cite{b2}, Painlev\'e analysis \cite{b3}, direct 
algebraic method \cite{b4}, tanh-function method 
\cite{b0}, \cite{b5}, \cite{b9}, sine-cosine 
method \cite{c10}, Darboux 
transformation \cite{d1, d2},  homotopy analysis 
method,  ($G'/G$)-expansion 
method  \cite{g0}, \cite{g3} and their extensions.

This paper addresses a 
modified  ($G'/G$)-expansion method 
to obtain new classes of solutions to nonlinear partial differential equations (NPDEs). This approach consists, first, in  linearizing 
the NPDE, and secondly, in  integrating the resulting partial differential equation (PDE) with a non-zero integration constant dependent constant.
Then,  the ($G'/G$)-expansion method is applied to deduce the solutions of the nonlinear PDE.

 The paper is organized as follows. In Section \eqref{sec1} we describe the  ($G'/G$)-method used to  solve  the NPDEs. 
We show that  any solution $\phi$ of the NPDE can be expanded  as polynomials depending on the new variable $G'/G$, where $G'$ is the derivative of $G$ with respect to the coordinate variables.  The general form of $\phi$ is also given. In Section \eqref{sec2} we explicitly solve the  Burgers partial differential equation, the Kadomtsev-Petviashvili  and KdV  equations. Relevant graphical representations are exhibited in each case. Section \eqref{sec3} is devoted to concluding remarks.

\section{Description of the method}\label{sec1}
The nonlinear partial differential equations (NPDEs), although difficult to solve,  are object of intensive  study in recent  literature as shown in  \cite{A5}-\cite{b8} and references therein. In this section, we consider the following types of NPDEs:

\begin{eqnarray}\label{1}
  Q(u, u_t, u_x, u_{xx}, \cdots)&=&0,\quad x,\,t\in\mathbb{R}, 
\end{eqnarray}

\begin{eqnarray}
\label{1eq}
 &P(u, u_t, u_x, u_y, u_{tt}, u_{xx}, u_{yy}, u_{tx}, u_{ty}, u_{xy}, \cdots)=0,\\
 &u_t=\frac{\partial u}{\partial t}, \, 
u_x=\frac{\partial u}{\partial x},\, u_{xx}=\frac{\partial^2 u}{\partial x^2},\,
 u_{tx}=\frac{\partial^2 u}{\partial t\partial x},\cdots; \,\,x,\,y,\,t\in\mathbb{R},
\end{eqnarray}

where $u=u(x, t)\in C^\infty(\mathbb{R}^2)$ and $v=v(x, y, t)\in C^\infty(\mathbb{R}^3)$ are solutions of (\ref{1}) and (\ref{1eq}), respectively.

Making  the transformations 

\begin{eqnarray}\label{2}
  u(x, t)=u(\nu )\quad \mbox{ and }\quad v(x, y, t)=v(\eta)
\end{eqnarray}

  with  

\begin{eqnarray}
\nu=x-\omega t,\; \eta =kx+\alpha y+\omega' t ,
\end{eqnarray}

where $\omega,\;\omega',\;k$ and $\alpha$ $\in\mathbb{R},$ the equations (\ref{1}) and (\ref{1eq}) can be rewritten as

\begin{eqnarray}\label{3}
  H(u, u', u'', \cdots)=0,
\end{eqnarray}

where 
$u(\xi)=u(\nu)$ for equation (\ref{1}) and $u(\xi)=v(\eta)$ for equation (\ref{1eq}).

Assume that equation (\ref{3}) is integrable 
with respect to
$\xi$ without cancelling the integrating constants. Then  introduce the following transformation \cite{b0} 
\begin{eqnarray}\label{4}
  u(\xi)=\phi (\xi)+c_1,\quad c_1\in\mathbb{R}.
\end{eqnarray}
Substituting (\ref{4}) into (\ref{3}) and setting the constant part equal to zero in the resulting nonlinear ODE in $\phi$ and assuming that the function $\phi$ and its derivatives have the following asymptotic values:
\begin{eqnarray}\label{condition1}
  \phi(\xi)\rightarrow \phi_\pm \quad \mbox{ as }\quad \xi\rightarrow \pm\infty,
\end{eqnarray}
for $n\geq 1$
\begin{eqnarray}\label{condition2}
  \phi^{(n)}\rightarrow 0\quad \mbox{ as }\quad \xi\rightarrow \pm\infty,
\end{eqnarray}
and  that $\phi_\pm$ satisfies the algebraic equation in $\phi$ \cite{b0}, we get the value of $c_1$.

The $(G\rq{}/G)$-expanding method consists in  setting
\begin{eqnarray}\label{6}
  \phi(\xi)=\alpha_m\Bigg(\frac{G'}{G} \Bigg)^m +\alpha_{m-1}\Bigg(\frac{G'}{G} \Bigg)^{m-1}+\cdots
\end{eqnarray}
 where $\alpha_m\neq 0$ and $G=G(\xi)$ satisfies the second order linear ordinary differential equation (LODE) in the form \cite{g2,b11}:
 \begin{eqnarray}\label{7}
   G''+\lambda G'+\mu G=0.
 \end{eqnarray}
The parameter $m$ can be found  by substituting along with equation  (\ref{7})  into equation (\ref{3})   and  considering a homogeneous balance between the highest order derivative and highest order nonlinear term in equation (\ref{3}), where $k,\;\alpha,\;\omega,\;a_0,\;a_1,\;\cdots,\;a_m$ are to be determined. Substituting (\ref{4}) and (\ref{6}) into (\ref{2})  yields a set of algebraic equations for $k,\;\alpha,\;a_0,\;a_1,\;\cdots,\;a_m$, vanishing all coefficients of $(G\rq{}/G)^i$. From these relations, $k,\;\alpha,\omega,\;\;a_0,\;a_1,\;\cdots,\;a_m$ can be obtained. Having determined these parameters, knowing that $m$ is a positive integer in most cases, and using (\ref{4}) and (\ref{6}) we obtain  analytical solutions $u( x , t )$ and  $v(x, y, t)$ of (\ref{1}) and (\ref{1eq}), respectively.\\

The solutions of the
equation (\ref{7}) are given  as follows:

$\bullet$ If $\lambda^2-4\mu> 0$, 
  \begin{eqnarray}
    G(\xi)=e^{-\frac{\lambda}{2}\xi}\left(k_1\cosh\Big(\frac{\sqrt{\lambda^2-4\mu}}{2}\xi\Big)+k_2\sinh\Big(\frac{\sqrt{\lambda^2-4\mu}}{2}\xi\Big)  \right)
\end{eqnarray}
and
\begin{eqnarray}
    \frac{G'}{G}=-\frac{\lambda}{2}+\frac{\sqrt{\lambda^2-4\mu}}{2}f(\xi;\lambda,\mu).
  \end{eqnarray}
where
\begin{eqnarray}
f(\xi;\lambda,\mu)=\frac{k_1\sinh\Big(\frac{\sqrt{\lambda^2-4\mu}}{2}\xi\Big)+k_2\cosh\Big(\frac{\sqrt{\lambda^2-4\mu}}{2}\xi\Big)}{k_1\cosh\Big(\frac{\sqrt{\lambda^2-4\mu}}{2}\xi\Big)+k_2\sinh\Big(\frac{\sqrt{\lambda^2-4\mu}}{2}\xi\Big)}
\end{eqnarray}

$\bullet$ If $\lambda^2-4\mu< 0$,
  \begin{eqnarray}
    G(\xi)=e^{-\frac{\lambda}{2}\xi}\left(k_1\cos\left(\frac{\sqrt{4\mu-\lambda^2}}{2}\xi\right)+k_2\sin\left(\frac{\sqrt{4\mu-\lambda^2}}{2}\xi\right)  \right)
\end{eqnarray}
and
\begin{eqnarray}
    \frac{G'}{G}=-\frac{\lambda}{2}+\frac{\sqrt{4\mu-\lambda^2}}{2}g(\xi;\lambda,\mu).
  \end{eqnarray}
where
\begin{eqnarray}
g(\xi;\lambda,\mu)=\frac{-k_1\sin\Big(\frac{\sqrt{4\mu-\lambda^2}}{2}\xi\Big)+k_2\cos\Big(\frac{\sqrt{4\mu-\lambda^2}}{2}\xi\Big)}{k_1\cos\Big(\frac{\sqrt{4\mu-\lambda^2}}{2}\xi\Big)+k_2\sin\Big(\frac{\sqrt{4\mu-\lambda^2}}{2}\xi\Big)},
\end{eqnarray}

$\bullet$ If $\lambda^2-4\mu=0$,  
  \begin{eqnarray}
    G(\xi)=(k_1\xi+k_2)e^{-\frac{\lambda}{2}\xi}\end{eqnarray}
and
\begin{eqnarray}
    \frac{G'}{G}=-\frac{\lambda}{2}+\frac{k_1}{k_1\xi+k_2}.
  \end{eqnarray} 
The equation (\ref{6}) can be re-expressed as 
\begin{eqnarray}
  \phi(\xi)&=&\sum_{k=1}^{m}\alpha_k\Bigg(\frac{G'(\xi)}{G(\xi)}\Bigg)^k+\alpha_0,\quad \alpha_0\in\mathbb{R},\quad\alpha_k\in\mathbb{R}\quad\forall k>0, 
\end{eqnarray}
giving
\begin{eqnarray}
\label{ralland:nomme}
  \phi'(\xi)&=&-\sum_{k=1}^{m} k \alpha_k \left\{\mu\Bigg(\frac{G'}{G}\Bigg)^{k-1}+\lambda \Bigg(\frac{G'}{G}\Bigg)^{k}+\Bigg(\frac{G'}{G}\Bigg)^{k+1} \right\}
\end{eqnarray}
and
\begin{eqnarray}
  \phi''(\xi)&=& \sum_{k=1}^m k\alpha_k \Bigg\{\mu^2(k-1)\Bigg(\frac{G'}{G}\Bigg)^{k-2}+\mu\lambda(2k-1)\Bigg(\frac{G'}{G}\Bigg)^{k-1}\\
            &+& k(\lambda^2+2\mu)\Bigg(\frac{G'}{G}\Bigg)^{k}+\lambda(2k+1)\Bigg(\frac{G'}{G}\Bigg)^{k+1}+(k+1)\Bigg(\frac{G'}{G}\Bigg)^{k+2} \Bigg\}\nonumber.
\end{eqnarray}

\section{Application to  NPDEs describing physical phenomena}\label{sec2}

In this section, we obtain new classes of  solutions  to some relevant NPDEs such as the
 Burgers equations \cite{b0}, KdV equation \cite{b0} and Kadomtsev-Petviashvili 
equation (also called KP equation in the literature) \cite{b1}. In each case the most relevant graphical representations are shown with an appropriate choice of parameters.


\subsection{  Burgers partial differential equation}
Let us consider the nonlinear Burgers equation in two-dimensional space-time given by the following relation :
\begin{eqnarray}\label{b}
  u_t+\alpha uu_x+\beta u_{xx}=0.
\end{eqnarray}
Using \eqref{2}, the equation (\ref{b}) takes the form
\begin{eqnarray}\label{c}
 M(u,u\rq{},u\rq{}\rq{})= -\omega u'+\alpha u u'+\beta u''=0.
\end{eqnarray}
Equation (\ref{c}) can be integrated to give the integral form
\begin{eqnarray}\label{o1}
 \int M(u,u\rq{},u\rq{}\rq{})\,d\xi= -\omega u+\frac{1}{2}\alpha u^2+\beta u'+c_1C=0,\,\,\, c_1C\in\mathbb{R}.
\end{eqnarray}
 Suppose that $u=\phi+c_1$. Then the
 equation \eqref{o1} can be transformed into the 
form
\begin{eqnarray}\label{eq}
 \alpha c_1\phi-\omega\phi+\frac{1}{2}\alpha\phi^2+\beta\phi'+c_1\left(\frac{1}{2}\alpha c_1+C-\omega\right)=0.
\end{eqnarray}
Using the equations (\ref{condition1}) and (\ref{condition2}), the functions $\phi_\pm$ satisfy the relation
\begin{eqnarray}
\alpha c_1\phi_\pm-\omega\phi_\pm+\frac{1}{2}\alpha\phi_\pm^2=0,
\end{eqnarray}
 with the condition
  \begin{eqnarray}
\label{rollandf}
c_1\left(\frac{1}{2}\alpha c_1+C-\omega\right)=0.
  \end{eqnarray}
 Two cases, deduced from the equation (\ref{rollandf}),  deserve investigation  in order 
to find explicit  solutions to the  equation \eqref{b}. 



\begin{proposition}[Case 1]  $c_1=0$ leads to the two following situations:

\begin{enumerate}
\item[(i)] $\xi=x-(2C-2\beta)t:$ We have

\begin{eqnarray}\label{p1}
 U_{11}=\left\{\begin{array}{ll}
\frac{2\beta}{\alpha}+\frac{2\beta}{\alpha} \left[ -\frac{\lambda}{2}+\frac{\sqrt{\lambda^2-4\mu}}{2}f(\xi;\lambda,\mu) \right],\mbox{ if } \,\lambda^2-4\mu>0,\\
\frac{2\beta}{\alpha}+\frac{2\beta}{\alpha} \left[-\frac{\lambda}{2}+\frac{\sqrt{4\mu-\lambda^2}}{2}g(\xi;\lambda,\mu)  \right],\,\mbox{ if }\lambda^2-4\mu<0\\
\frac{2\beta}{\alpha}+\frac{2\beta}{\alpha} \left(-\frac{\lambda}{2}+\frac{k_1}{k_1\xi+k_2}  \right), \mbox{ if }\,\lambda^2-4\mu=0,
\end{array}\right.
\end{eqnarray}


\item[(ii)]
 $\xi=x-(2C-\beta\mu)t:$ We get

\begin{eqnarray}\label{p4}
 U_{12}=\left\{\begin{array}{ll}
\frac{2\beta\mu}{\alpha}+\frac{2\beta}{\alpha} \left[ -\frac{\lambda}{2}+\frac{\sqrt{\lambda^2-4\mu}}{2}f(\xi;\lambda,\mu) \right],\mbox{ if }\, \lambda^2-4\mu>0,\\
\frac{2\beta\mu}{\alpha}+\frac{2\beta}{\alpha} \left[-\frac{\lambda}{2}+\frac{\sqrt{4\mu-\lambda^2}}{2}g(\xi;\lambda,\mu) \right] \mbox{ if } \,\lambda^2-4\mu<0\\
\frac{2\beta\mu}{\alpha}+\frac{2\beta}{\alpha} \left[-\frac{\lambda}{2}+\frac{k_1}{k_1\xi+k_2}  \right] \mbox{ if }   \lambda^2-4\mu=0,
\end{array}\right.
\end{eqnarray}
\end{enumerate}
\end{proposition}


%

\begin{figure}[!h]
\begin{minipage}[t]{.45\linewidth}
  \begin{center}
  \includegraphics[scale=.35]{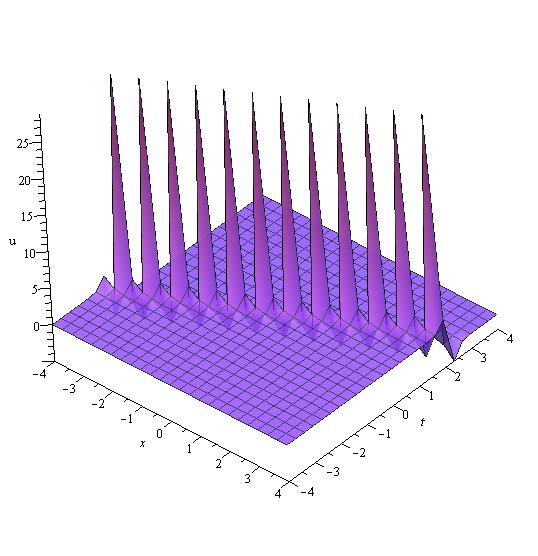}
  \end{center}
  \caption{Solution $U_{11}$ for $\alpha=2,\,\beta=1,\,\mu=4,\,c_1=2,\;c_2=2,\,C=2$}\label{fig_1}
\end{minipage}
\hfill
\begin{minipage}[t]{.45\linewidth}
 \begin{center}
  \includegraphics[scale=.35]{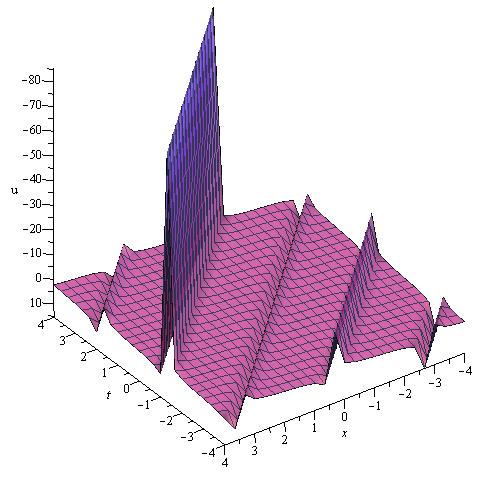}
  \end{center}
\caption{Solution $U_{12}$ for $\alpha=2,\,\beta=1,\,\mu=5,\,c_1=2,\,c_2=4,\,C=2$}
\end{minipage}
\end{figure}

{\bf Proof of relations (\ref{p1}) and
 (\ref{p4})}.

Equation (\ref{eq}) becomes
\begin{eqnarray}\label{eq1}
  -\omega\phi+\frac{1}{2}\alpha\phi^2+\beta\phi'=0.
\end{eqnarray}
Using 
 (\ref{6}) and  balancing $\phi^2$ into $\phi'$, we get 
\begin{eqnarray}
2m=m+1\quad \mbox{ so},\quad  m=1.
\end{eqnarray}
 Finally
\begin{eqnarray}
\label{rolland:fik}
\phi(\xi)=\alpha_1\Big(\frac{G'}{G} \Big)+\alpha_0.
\end{eqnarray}
Substituting  (\ref{rolland:fik}) and (\ref{ralland:nomme}) (for $m=1$)
in  (\ref{eq1}), we obtain the following expression

\begin{eqnarray}
\label{rolland:finangnon}
  -\omega\alpha_1\Big(\frac{G'}{G} \Big)-\omega\alpha_0+\frac{1}{2}\alpha\alpha_0^2+\alpha \alpha_0\alpha_1\Big(\frac{G'}{G} \Big)+\frac{1}{2}\alpha \alpha_1^2\Big(\frac{G'}{G} \Big)^2\cr
+\beta\alpha _0-\beta\alpha_1\mu-\beta\alpha_1\mu\Big(\frac{G'}{G} \Big)-\beta\alpha_1\Big(\frac{G'}{G} \Big)^2=0.
\end{eqnarray} 

Vanishing the polynomial coefficients yields  the system

\begin{eqnarray}
\left\{\begin{array}{ll}
-\omega\alpha_0+\frac{1}{2}\alpha\alpha_0^2+\beta\alpha_0-\beta\alpha_1\mu&=0\cr
-\omega\alpha_1+\alpha\alpha_0\alpha_1-\beta\alpha_1\mu&=0\cr
\frac{1}{2}\alpha\alpha_1^2-\beta\alpha_1&=0
\end{array}\right.
\end{eqnarray}

whose solutions are given by 

\begin{eqnarray}\label{ddd}
\left\{\begin{array}{ll}
\omega&=2C-2\beta\cr
\alpha_0&=\frac{2\beta}{\alpha}\cr
\alpha_1&=\frac{2\beta}{\alpha},
\end{array}\right. \quad\quad\mbox{ or }\quad\quad
\left\{\begin{array}{ll}
\omega&=2C-\beta\mu\cr
\alpha_0&=\frac{2\beta\mu}{\alpha}\cr
\alpha_1&=\frac{2\beta}{\alpha}.
\end{array}\right.
\end{eqnarray}
The relations (\ref{p1}) and
 (\ref{p4})
 are therefore well satisfied.$\blacksquare$


\begin{proposition}[Case 2] $c_1=\frac{2}{\alpha}(\omega-C)$ furnishes two situations:
\begin{enumerate}
\item[(i)]  $\xi=x-(2C-2\beta+\beta\mu)t$

\begin{eqnarray}\label{pp1}
 U_{13}=\left\{\begin{array}{ll}
\frac{-2\beta-\beta\lambda+2C+2\beta\mu}{\alpha}+\frac{\beta\sqrt{\lambda^2-4\mu}}{\alpha}f(\xi;\lambda,\mu), \mbox{ if }\,\lambda^2-4\mu>0,\\
\frac{-2\beta-\beta\lambda+2C+2\beta\mu}{\alpha}+\frac{\beta\sqrt{4\mu-\lambda^2}}{\alpha}g(\xi;\lambda,\mu),\mbox{ if }\,\lambda^2-4\mu<0,\\
 \frac{-2\beta+2C-\beta\lambda+2\beta\mu}{\alpha}+\frac{2\beta}{\alpha}\left(\frac{k_1}{k_1\xi+k_2}\right),\mbox{ if } \lambda^2-4\mu=0.
\end{array}\right.
\end{eqnarray}


\item[(ii)]   $\xi=x-(2C-\beta\mu)t:$

\begin{eqnarray}\label{pp4}
 U_{14}=\left\{\begin{array}{ll}
\frac{-\beta\lambda+2C}{\alpha}+\frac{\beta\sqrt{\lambda^2-4\mu}}{\alpha}f(\xi;\lambda,\mu), \mbox{ if } \,\lambda^2-4\mu>0,\\
\frac{-\beta\lambda+2C}{\alpha}+\frac{\beta\sqrt{4\mu-\lambda^2}}{\alpha}g(\xi;\lambda,\mu),\mbox{ if }\, \lambda^2-4\mu<0,\\
\frac{-\beta\lambda+2C}{\alpha}+\frac{2\beta}{\alpha}\left(\frac{k_1}{k_1\xi+k_2}\right)\mbox{ if }\,  \lambda^2-4\mu=0.
\end{array}\right.
\end{eqnarray}

%
\begin{figure}[!h]
 \begin{minipage}[t]{.45\linewidth}
   \begin{center}
  \includegraphics[scale=.3]{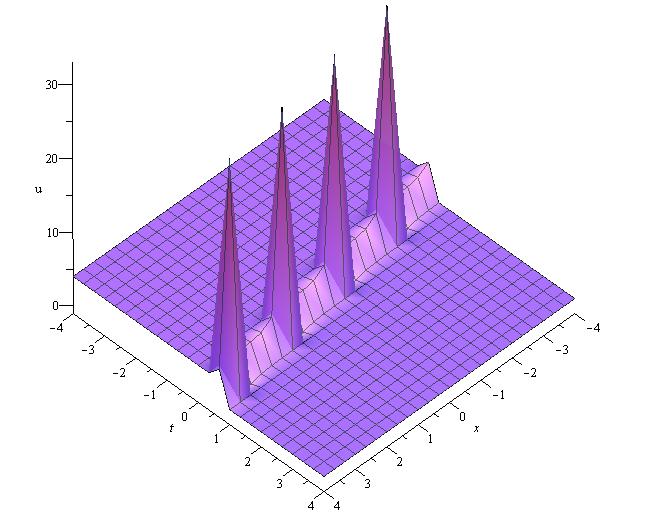}
  \end{center}
  \caption{Solution $U_{13}$ for $\alpha=2,\,\beta=1,\,\lambda=5,\,\mu=4,\,c_1=2,\;c_2=4,\,C=2$}\label{fig_3}
\end{minipage}
\hfill
\begin{minipage}[t]{.45\linewidth}
 \begin{center}
  \includegraphics[scale=.38]{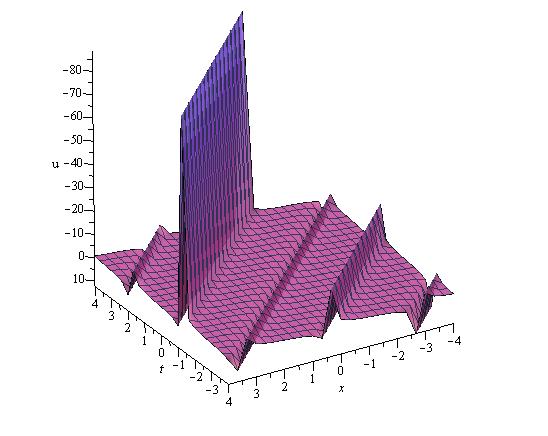}
  \end{center}
    \caption{Solution $U_{14}$ for $\alpha=2,\,\beta=1,\,\lambda=4,\,\mu=5,\,c_1=2,\;c_2=4,\,C=2$}\label{fig_4}
\end{minipage}
\end{figure}
\end{enumerate}
\end{proposition}
{\bf Proof of relations (\ref{pp1}) and (\ref{pp4}))}.

 The equation (\ref{eq}) becomes
\begin{eqnarray}\label{eq2}
  \omega\phi-2C\phi+\frac{1}{2}\alpha\phi^2+\beta\phi'=0.
\end{eqnarray}
Applying $(G\rq{}/G)$-expansion method on (\ref{eq2}),  using (\ref{6}) 
and  balancing $\phi^2$ into $\phi'$, we get $m=1$ and 
\beqar
\label{tttt}
\phi(\xi)=\alpha_1\big(\frac{G'}{G} \big)+\alpha_0.
\eeqar
Using 
(\ref{tttt}), the relation (\ref{eq2}) is re-expressed  as a polynomial in 
$(G\rq{}/G)$ and we obtain
\begin{eqnarray}
  \omega\alpha_1\Big(\frac{G'}{G} \Big)+\omega\alpha_0-2C\alpha_1\Big(\frac{G'}{G} \Big)-2C\alpha_0+\frac{1}{2}\alpha\alpha_0^2+\alpha \alpha_0\alpha_1\Big(\frac{G'}{G} \Big)\cr
  +\frac{1}{2}\alpha \alpha_1^2\Big(\frac{G'}{G} \Big)^2+\beta\alpha _0-\beta\alpha_1\mu-\beta\alpha_1\mu\Big(\frac{G'}{G} \Big)^2-\beta\alpha_1\Big(\frac{G'}{G} \Big)^2=0.
\end{eqnarray} 
Vanishing the polynomial coefficients provides 
 the system
\begin{eqnarray}
\left\{\begin{array}{ll}
\omega\alpha_0-2C\alpha_0+\frac{1}{2}\alpha\alpha_0^2+\beta\alpha_0-\beta\alpha_1\mu&=0\\
\omega\alpha_1-2C\alpha_1+\alpha\alpha_0\alpha_1-\beta\alpha_1\mu&=0\\
\frac{1}{2}\alpha\alpha_1^2-\beta\alpha_1&=0
\end{array}\right.
\end{eqnarray}
which can be solved to give the solutions
\begin{eqnarray}\label{lllll}
 \left\{\begin{array}{ll}
\omega&=2C-2\beta+\beta\mu\\
\alpha_0&=\frac{2\beta}{\alpha}\\
\alpha_1&=\frac{2\beta}{\alpha},
\end{array}\right. \quad\mbox{or}\quad
 \left\{\begin{array}{ll}
\omega&=2C-\beta\mu\\
\alpha_0&=\frac{2\beta\mu}{\alpha}\\
\alpha_1&=\frac{2\beta}{\alpha}.
\end{array}\right.
\end{eqnarray}
Similarly as in  the previous  case, the relations (\ref{pp1}) and (\ref{pp4}) are well satisfied.$\blacksquare$


\subsection{KdV nonlinear equation}
We consider  the following  KdV equation:
\begin{eqnarray}\label{30}
 u_t+\alpha u u_x+\gamma u_{xxx}=0.
\end{eqnarray}
To solve  it we use the transformation (\ref{2}) to reduce the equation (\ref{30}) into  an ordinary differential equation (ODE):
\begin{eqnarray}\label{31}
  -\omega u+\frac{1}{2}\alpha u^2+\gamma u''+c_1C=0,
\end{eqnarray}
where $c_1C$ is constant. Combining (\ref{4}) and
(\ref{31}), we get
\begin{eqnarray}\label{32}
  (\alpha c_1-\omega)\phi+\frac{1}{2}\alpha\phi^2+\gamma\phi''+c_1(\frac{1}{2}\alpha c_1+C-\omega)=0.
\end{eqnarray}
The solutions of \eqref{30} are also given in two different  cases.

\begin{proposition}[Case 1] $c_1=0$ leads to two situations:
\begin{enumerate}
\item[(i)] If  $\xi=x-\gamma(4\mu-\lambda^2)t,$  we get

\begin{eqnarray}\label{R1}
 U_{21}=\left\{\begin{array}{ll}
\frac{-4\gamma\mu+\gamma\lambda^2}{\alpha}-\frac{3\gamma(\lambda^2-4\mu)}{\alpha} f^2(\xi;\lambda;\mu) , \mbox{ if }\,\lambda^2-4\mu>0,\\
\frac{-4\gamma\mu+\gamma\lambda^2}{\alpha}-\frac{3\gamma(\lambda^2-4\mu)}{\alpha}g^2(\xi;\lambda;\mu), \mbox{ if }\, \lambda^2-4\mu<0,\\
\frac{-4\gamma\mu+6\gamma\mu\lambda-5\gamma\lambda^2}{\alpha}+\frac{12\gamma(\lambda-\mu)}{\alpha}\left(\frac{k_1}{k_1\xi+k_2}\right)-\frac{12\gamma}{\alpha}\left(\frac{k_1}{k_1\xi+k_2}\right)^2,
\mbox{ if } \lambda^2-4\mu=0.
\end{array}\right.
\end{eqnarray}

\item[(ii)]   If  $\xi=x+\gamma(4\mu-\lambda^2)t$, we have:

\begin{eqnarray}\label{R4}
 U_{22}=\left\{\begin{array}{ll}
\frac{6\gamma(\lambda-\mu)\sqrt{\lambda^2-4\mu}}{\alpha} f(\xi;\lambda,\mu)
-\frac{3\gamma(\lambda^2-4\mu)}{\alpha} f^2(\xi;\lambda;\mu)\\ +\frac{-12\gamma\mu+6\gamma\mu\lambda-3\gamma\lambda^2}{\alpha}, \mbox{ if }\lambda^2-4\mu> 0,\\
\frac{6\gamma(\lambda-\mu)\sqrt{\lambda^2-4\mu}}{\alpha} g(\xi;\lambda,\mu)
-\frac{3\gamma(\lambda^2-4\mu)}{\alpha}g^2(\xi;\lambda;\mu)\\
+\frac{-12\gamma\mu+
6\gamma\mu\lambda-3\gamma\lambda^2}{\alpha}, 
\mbox{ if }\, \lambda^2-4\mu< 0,\\
\frac{-12\gamma\mu+6\gamma\mu\lambda-3\gamma\lambda^2}{\alpha}+\frac{12\gamma(\lambda-\mu)}{\alpha}\left(\frac{k_1}{k_1\xi+k_2}\right)\\
-\frac{12\gamma}{\alpha}\left(\frac{k_1}{k_1\xi+k_2}\right)^2, \mbox{ if }\,\lambda^2-4\mu= 0.
\end{array}\right.
\end{eqnarray}

\end{enumerate}
\end{proposition}
\begin{figure}[!h]
\begin{minipage}[t]{.45\linewidth}
  \begin{center}
  \includegraphics[scale=.37]{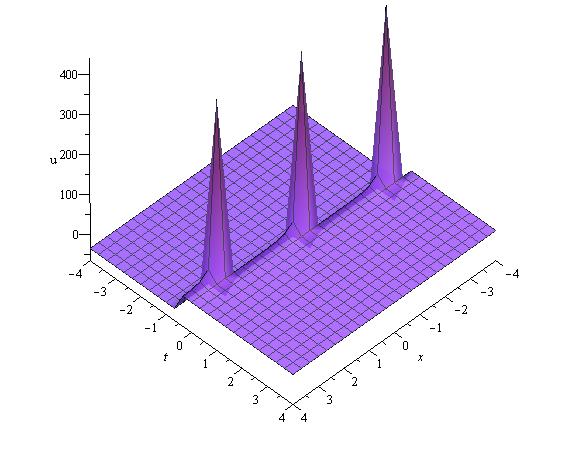}
  \end{center}
    \caption{Solution $U_{21}$ for $\alpha=2,\,\gamma=1,\,\lambda=5,\,\mu=4,\,c_1=2,\;c_2=4,\,C=$}\label{fig_5}
\end{minipage}
\hfill
\begin{minipage}[t]{.45\linewidth}
 \begin{center}
  \includegraphics[scale=.33]{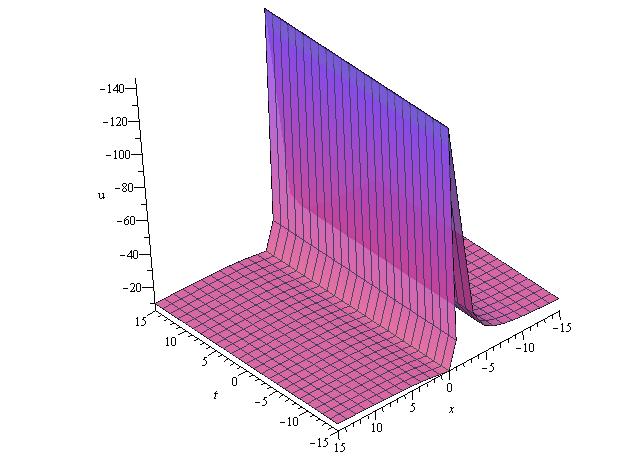}
  \end{center}
 \caption{Solution of $U_{22}$ for $\alpha=2,\,\gamma=1,\,\lambda=4,\,\mu=5,\,c_1=2,\;c_2=2,\,C=4$}\label{fig_5}
\end{minipage}
\end{figure}

{\bf Proof of relations (\ref{R1}) and (\ref{R4})}.

  The equation (\ref{32}) is reduced to
\begin{eqnarray}\label{33}
  -\omega\phi+\frac{1}{2}\alpha\phi^2+\gamma\phi''=0.
\end{eqnarray}
Taking into account (\ref{6}), and balancing $\phi^2$ into $\phi''$, we get $m=2$ and arrive at the following system: 
\begin{eqnarray}
\label{ytem}
\left\{\begin{array}{ll}
-\omega\alpha_0+\frac{1}{2}\alpha\alpha_0^2+2\gamma\alpha_2\mu^2+\gamma\alpha_1\lambda\mu&=0\\
-\omega\alpha_1+\alpha\alpha_0\alpha_1+6\gamma\alpha_2\lambda\mu+2\gamma\alpha_1\mu+
\alpha_1\mu+\alpha_1\lambda^2\gamma&=0\\
-\omega\alpha_2+\alpha\alpha_0\alpha_2+\frac{1}{2}\alpha\alpha_1^2+4\gamma\alpha_2\lambda^2+3\alpha_1\gamma\lambda+8\alpha_2\mu\gamma&=0\\
\alpha\alpha_1\alpha_2+10\gamma\alpha_2\lambda+2\alpha_1\gamma&=0\\
\frac{1}{2}\alpha\alpha_2+6\alpha_2\gamma&=0.
\end{array}\right.
\end{eqnarray}
which can be  solved to   give
\begin{eqnarray}\label{tt}
\left\{\begin{array}{ll}
\omega&=\gamma(4\mu-\lambda^2)\\
\alpha_0&=\frac{-2\gamma(2\mu+\lambda^2)}{\alpha}\\
\alpha_1&=-\frac{12\gamma\lambda}{\alpha}\\
\alpha_2&=-\frac{12\gamma}{\alpha}
\end{array}\right. \quad\mbox{or}\quad
 \left\{\begin{array}{ll}
\omega&=-\gamma(4\mu-\lambda^2)\\
\alpha_0&=\frac{-12\gamma\mu}{\alpha}\\
\alpha_1&=-\frac{12\gamma\lambda}{\alpha}\\
\alpha_2&=-\frac{12\gamma}{\alpha}.
\end{array}\right.
\end{eqnarray}
The relations (\ref{R1}) and (\ref{R4}) then hold.$\blacksquare$
%
%


\begin{proposition} [Case 2] $c_1=\frac{2\omega-2C}{\alpha}$ amounts to the  solutions of \eqref{30}  in the  two following situations:
\begin{enumerate}
\item[(i)] $\xi=x-(2C+4\gamma\mu-\lambda^2\gamma)t$ gives

\begin{eqnarray}\label{RR1}
 U_{23}=\left\{\begin{array}{ll}
\frac{-4\gamma\mu+\gamma\lambda^2+2C}{\alpha}-
\frac{3\gamma(\lambda^2-4\mu)}{\alpha} f^2 (\xi;\lambda;\mu),\mbox{ if }\,\lambda^2-4\mu>0,\\
\frac{-4\gamma\mu+\gamma\lambda^2+2C}{\alpha}-
\frac{3\gamma(\lambda^2-4\mu)}{\alpha}g^2(\xi;\lambda;\mu),\mbox{ if }\,\lambda^2-4\mu<0\mbox{ and }\\
\frac{-4\gamma\mu+\gamma\lambda^2+2C}{\alpha}-\frac{12\gamma}{\alpha}\left(\frac{k_1}{k_1\xi+k_2}\right)^2,\mbox{ if }\,\lambda^2-4\mu=0.
\end{array}\right.
\end{eqnarray}


\item[(ii)]   $\xi=x-(2C+4\gamma\mu+\lambda^2\gamma)t$ yields

\begin{eqnarray}\label{RR4}
 U_{24}=\left\{\begin{array}{ll}
\frac{-12\gamma\mu+3\gamma\lambda^2+2C}{\alpha}-
\frac{3\gamma(\lambda^2-4\mu)}{\alpha} f^2(\xi;\lambda;\mu),\mbox{ if }\,\lambda^2- 4\mu>0,  \\
\frac{-12\gamma\mu+3\gamma\lambda^2+2C}{\alpha}-\frac{3\gamma(\lambda^2-4\mu)}{\alpha}g^2 (\xi;\lambda;\mu),\mbox{ if }\, \lambda^2-4\mu<0,\\
\frac{-12\gamma\mu+3\gamma\lambda^2+2C}{\alpha}-\frac{12\gamma}{\alpha}\left(\frac{k_1}{k_1\xi+k_2}\right)^2,\mbox{ if }\, \lambda^2-4\mu=0.
\end{array}\right.
\end{eqnarray}

\end{enumerate}
\end{proposition}
\begin{figure}[!h]
\begin{minipage}[t]{.45\linewidth}
  \begin{center}
  \includegraphics[scale=.34]{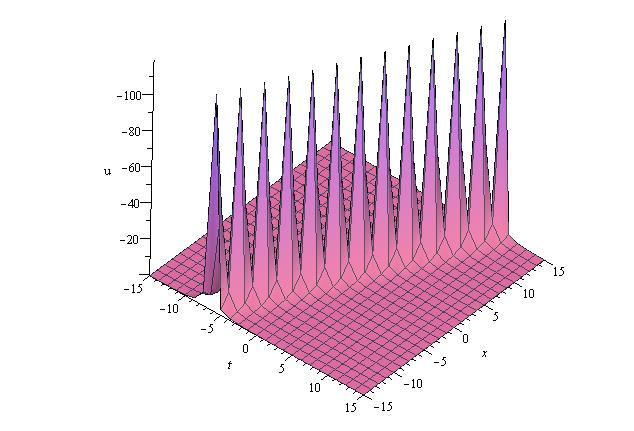}
  \end{center}
  \caption{Solution of $U_{23}$ for $\alpha=2,\,\gamma=5,\,\lambda=1,\,\mu=1/4,\,c_1=2,\;c_2=4,\,C=1,\,\beta=1$}\label{fig_7}
\end{minipage}
\hfill
\begin{minipage}[t]{.4\linewidth}
 \begin{center}
  \includegraphics[scale=.25]{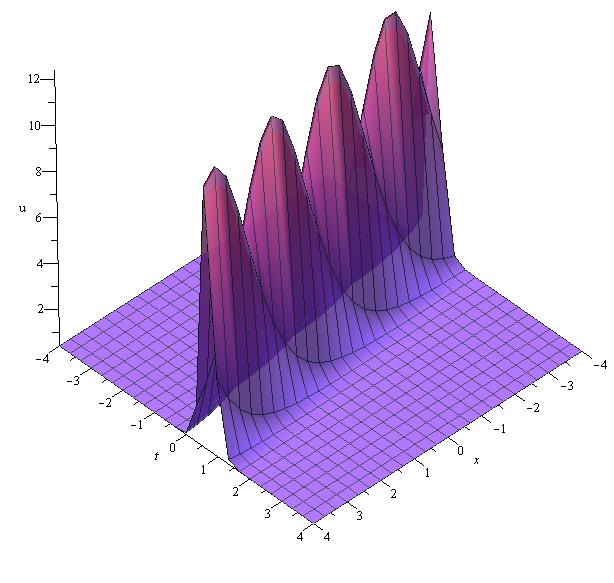}
  \end{center}
  \caption{Soliton solution of $U_{24}$ for $\alpha=1/2,\,\gamma=1,\,\lambda=2,\,\mu=1/2,\,c_1=1,\;c_2=0.3,\,C=0.1$}\label{fig_8}
\end{minipage}
\end{figure}


%
{\bf Proof of relations (\ref{RR1}) and (\ref{RR4})}.

The equation (\ref{32}) can be re-expressed as
\begin{eqnarray}\label{37}
  (\omega-2C)\phi+\frac{1}{2}\alpha\phi^2+\gamma\phi''=0.
\end{eqnarray}
Using the equation  (\ref{6}) and balancing $\phi^2$ into $\phi''$  imply that  $m=2$ and we arrive at the following system:
\begin{eqnarray}
\left\{\begin{array}{ll}
\omega\alpha_0-2C\alpha_0+\frac{1}{2}\alpha\alpha_0^2+2\gamma\alpha_2\mu^2+\gamma\alpha_1\lambda\mu^2&=0\\
\omega\alpha_1-2C\alpha_1+\alpha\alpha_0\alpha_1+6\gamma\alpha_2\lambda
\mu+2\gamma\alpha_1\mu+\alpha_1\mu+\alpha_1\lambda^2\gamma&=0\\
\omega\alpha_2-2C\alpha_2+\alpha\alpha_0\alpha_2+\frac{1}{2}\alpha\alpha_1^2+4\gamma\alpha_2\lambda^2+3\alpha_1\gamma\lambda+8\alpha_2\mu\gamma&=0\\
\alpha\alpha_1\alpha_2+10\gamma\alpha_2\lambda+2\alpha_1\gamma&=0\\
\frac{1}{2}\alpha\alpha_2^2+6\alpha_2\gamma&=0
\end{array}\right.
\end{eqnarray}
whose solutions are given by
\begin{eqnarray}\label{xx}
\left\{\begin{array}{ll}
\omega&=2C+4\gamma\mu-\lambda^2\gamma\\
\alpha_0&=-\frac{12\gamma\mu}{\alpha}\\
\alpha_1&=-\frac{12\gamma\lambda}{\alpha}\\
\alpha_2&=-\frac{12\gamma}{\alpha}
\end{array}\right.
\quad\mbox{or}\quad
\left\{\begin{array}{ll}
\omega&=2C-4\gamma\mu+\lambda^2\gamma\\
\alpha_0&=\frac{2\gamma(2\mu+\lambda^2)}{\alpha}\\
\alpha_1&=-\frac{12\gamma\lambda}{\alpha}\\
\alpha_2&=-\frac{12\gamma}{\alpha}.
\end{array}\right.
\end{eqnarray}
Therefore we simply get relations (\ref{RR1}) and (\ref{RR4})). $\blacksquare$

\subsection{Kadomtsev-Petviashvili nonlinear equation and its solutions }
This subsection is  devoted to the resolution of the Kadomtsev-Petviashvili NPDE in three dimensions given by
 \begin{eqnarray}
\label{aa}
(u_t+6uu_x+u_{xxx})_x+3\sigma^2u_{yy}=0,\;\;\;\; \sigma^2=\pm1.
\end{eqnarray}
Using the transformation (\ref{2}), the equation (\ref{aa}) takes the form 
\begin{eqnarray}\label{bb}
  N(u,u',u'',u''')=k \omega u''+6k^2(u'u)'+k^4u^{(4)}+3\sigma^2\alpha^2u''=0.
\end{eqnarray}
Integrating this latter equation,
we get the following
 nonlinear ODE 
\begin{eqnarray}
\label{eee}
\int \, N(u,u',u'',u''')\,d\xi=(k\omega+3\sigma^2\alpha^2)u'+6k^2u'u+k^4u'''+c_1C=0,
\end{eqnarray}
where $c_1C$ is a constant to be determined.   Substituting
(\ref{4}) into (\ref{eee}), we simply get
\begin{eqnarray}\label{bbb}
  (k\omega +3\sigma^2 \alpha^2+6c_1k^2)\phi+3k^2\phi^2+k^4\phi''
+c_1(k\omega+3\sigma^2\alpha^2+3k^2c_1+C)=0.
\end{eqnarray}
The equation \eqref{aa} can be now solved in different cases to give the next results:

\begin{proposition}[Case 1]  $c_1=0$.
The  solutions of (\ref{aa}) are:
\begin{enumerate}
\item[(1)] If  $\xi=x-(4k^4\mu-3\sigma^2\alpha^2-k^4\lambda^2)t/k:$

\begin{eqnarray}\label{f1}
 U_{31}=\left\{\begin{array}{ll}
-2k^2\mu+\frac{1}{2}k^2\lambda^2-\frac{k^2(\lambda^2-4\mu)}{2} f(\xi;\lambda,\mu)^2 ,\,\, \mbox{   if  }\,\, \lambda^2-4\mu> 0, \\
-2k^2\mu+\frac{1}{2}k^2\lambda^2-\frac{k^2(4\mu-\lambda^2)}{2}g(\xi;\lambda,\mu)^2,\,\, \mbox{ if  } \lambda^2-4\mu< 0,\\
 -2k^2\mu+\frac{1}{2}k^2\lambda^2-2k^2\left(\frac{c_1}{c_1\xi+C_2}\right)^2,\,\, \mbox{ if  } \lambda^2-4\mu= 0.
\end{array}\right.
\end{eqnarray}

\item[(ii)] If $\xi=x-(4k^4\mu-3\sigma^2\alpha^2-k^4\lambda^2)t/k:$

\begin{eqnarray}\label{f4}
 U_{32}=\left\{\begin{array}{ll}
-\frac{2}{3}k^2\mu+\frac{1}{6}k^2\lambda^2-\frac{k^2(\lambda^2-4\mu)}{2} f(\xi;\lambda,\mu)^2 ,\,\,\mbox{  if  } \lambda^2-4\mu> 0, \\
-\frac{2}{3}k^2\mu+\frac{1}{6}k^2\lambda^2-\frac{k^2(4\mu-\lambda^2)}{2}g(\xi;\lambda,\mu)^2,\,\,\mbox{ if } \lambda^2-4\mu< 0,\\
-\frac{2}{3}k^2\mu+\frac{1}{6}k^2\lambda^2-2k^2\left(\frac{c_1}{c_1\xi+C_2}\right)^2,\,\,\mbox{   if   } \lambda^2-4\mu= 0.
\end{array}\right.
\end{eqnarray}

\end{enumerate}
\end{proposition}
{\bf Proof
 of relations \eqref{f1} and \eqref{f4}}.  

The   equation (\ref{bbb}) takes the form
\begin{eqnarray}
 (k\omega +3\sigma^2 \alpha^2)\phi+3k^2\phi^2+k^4\phi''=0.
\end{eqnarray}
Let us consider  (\ref{6}) and balance $\phi^2$ into $\phi''$ to get $m=2$.Then  we arrive at the following 
system
\begin{eqnarray}
\label{systemss}
\left\{\begin{array}{ll}
3k^2\alpha_2^2+6k^4\alpha_2&=0\\
6k^4\alpha\alpha_2+10k^4\alpha_2\lambda+2k^4\alpha_1&=0\\
3\sigma^2\alpha^2\alpha_2+k\omega\alpha_2+3k^2\alpha_1^2+3k^4\alpha_1\lambda 
+8k^4\alpha_2\mu+6k^2\alpha_0\alpha_1\alpha_2+4k^4\alpha_2\lambda^2&=0\\
k\omega\alpha_1+2k^4\alpha_1\mu+\alpha k^2\alpha_0\alpha_1+3\sigma^2\alpha^2\alpha_1 
+6k^4\alpha_2\lambda\mu+k^4\alpha_1\lambda^2&=0\\
k\omega\alpha_0+k^4\alpha_1\lambda\mu+3k^2\alpha_0^2+3\sigma^2\alpha^2\alpha_0+2k^4\alpha_2\mu^2&=0
\end{array}\right.
\end{eqnarray}
giving the solutions 
\begin{eqnarray}\label{hh}
\left\{\begin{array}{l}
\omega=\frac{4k^4\mu-3\sigma^2\alpha^2-k^4\lambda^2}{k}\\
\alpha_0=-2k^2\mu\\
\alpha_1=-2k^2\mu\\
\alpha_2=-2k^2
\end{array}\right.\quad\mbox{ or }\quad
\left\{\begin{array}{l}
\omega=-\frac{4k^4\mu+3\sigma^2\alpha^2-k^4\lambda^2}{k}\\
\alpha_0=-\frac{2}{3}k^2\mu-\frac{1}{3}k^2\lambda^2\\
\alpha_1=-2k^2\lambda\\
\alpha_2=-2k^2.
\end{array}\right.
\end{eqnarray}
There result the relations  \eqref{f1} and \eqref{f4}. $\blacksquare$

\begin{proposition} [Case 2]  $c_1=\frac{-C-3\sigma^2\alpha^2-k^2\omega}{3}.$ It leads to 
the  solutions of (\ref{aa}) 
in two situations:
\begin{enumerate}
\item[(i)]  $ \xi=kx+\alpha 
y-(4k^4\mu-2k^2C-k^4\lambda^2+6k^2
\sigma^2\alpha^2-3\sigma^2\alpha^2)
t/k(-1+2k^3)$ provides

\begin{eqnarray}\label{q1}
 U_{33}=\left\{\begin{array}{ll}
\frac{-12k^2\mu+3k^2\lambda^2-2C-6\sigma^2
\alpha^2-2k^2\omega}{6}-\frac{k^2(\lambda^2
-4\mu)}{2}f(\xi;\lambda,\mu)^2,\\
\mbox{ if }   \lambda^2-4\mu> 0,\\
-\frac{-12k^2\mu+3k^2\lambda^2-2C-6\sigma^2
\alpha^2-2k^2\omega}{6} -\frac{k^2(4\mu-\lambda^2)}{2}g(\xi;\lambda,
\mu)^2 \\
 \mbox{ if } \lambda^2-4\mu< 0,\\
\frac{-12k^2\mu+3k^2\lambda^2-2C-6\sigma^2
\alpha^2-2k^2\omega}{6} -2k^2\Big(\frac{c_1}{c_1
\xi+C_2}\Big)^2 \\
\mbox{ if } \lambda^2-4\mu= 0
\end{array}\right.
\end{eqnarray}


\item[(ii)]  $\xi=k x+\alpha y-(-4k^4\mu+2k^2C+k^4\lambda^2+6k^2\sigma^2
\alpha^2-3\sigma^2\alpha^2)t/k(-1+2k^3)$ gives

\begin{eqnarray}\label{q4}
 U_{34}=\left\{\begin{array}{ll}
\frac{-4k^2\mu+k^2\lambda^2-2C-6\sigma^2
\alpha^2-2k^2\omega}{6}-\frac{k^2(\lambda^2-4
\mu)}{2}f(\xi;\lambda,\mu)^2, \\
\mbox{ if } \lambda^2-4\mu> 0,\\
\frac{-4k^2\mu+k^2\lambda^2-2C-6\sigma^2
\alpha^2-2k^2\omega}{6}-\frac{k^2(4\mu-
\lambda^2)}{2}g(\xi;\lambda,\mu)^2, \\
\mbox{ if } \lambda^2-4\mu< 0\\
 \frac{-4k^2\mu+k^2\lambda^2-2C-6\sigma^2
\alpha^2-2k^2\omega}{6}-2k^2\Big(\frac{c_1}{c_1
\xi+C_2}\Big)^2\\
\mbox{ if } \lambda^2-4\mu= 0.
\end{array}\right.
\end{eqnarray}

\end{enumerate}
\end{proposition}

 
{\bf Proof of \eqref{q1} and \eqref{q4}}. 

The equation 
(\ref{bbb}) can be rewritten as
\begin{eqnarray}\label{377}
  (k\omega+3\sigma^2\alpha^2-2k^2C-6k^2\sigma^2
\alpha^2-2k^4\omega)\phi+3k^2\phi^2+k^4\phi''=0.
\end{eqnarray}
Considering (\ref{6}) and balancing $\phi^2$ into $\phi''$, we get $m=2$ and
obtain the following 
system
\begin{eqnarray}\label{www}
\left\{\begin{array}{ll}
3k^2C_2^2+6k^4\alpha_2&=0\\
6k^2\alpha_1\alpha_2+10k^4\alpha_2\lambda+2k^4\alpha_1&=0\\
-6k^2\sigma^2\alpha^2\alpha_2-2k^2\omega\alpha_2+
6k^2\alpha_0\alpha_2-2k^2C\alpha_2+3k^4\alpha_1\lambda\\
+3k^2\alpha_1^2+k\omega\alpha_2+8k^4\alpha_2\mu+
3\sigma^2\alpha^2\alpha_2+4k^4\alpha_2\lambda^2&=0\\
2k^4\alpha_1\mu-2k^4\omega\alpha_1+2k^4\alpha_1
\lambda^2+k\omega\alpha_1-2k^2C\alpha_1+6k^4\alpha_2\\
-6k^2\sigma^2\alpha^2\alpha_1+6k^2\alpha_0
\alpha_1+3\sigma^2\alpha^2\alpha_1&=0\\
k\omega\alpha_0-2k^2C\alpha_0-
6k^2\sigma^2\alpha^2\alpha_0+3k^2\alpha_0^2\\
+3\sigma^2\alpha^2\alpha_0+2k^4\alpha_2
\mu^2-2k^4\omega\alpha_0+k^4\alpha_1\lambda\mu&=0
\end{array}\right.
\end{eqnarray}
Its  solutions are given by
\begin{eqnarray}
&\left\{\begin{array}{ll}
\omega=-\frac{2k^2+6k^2\sigma^2\alpha^2+
4k^4\mu-3\sigma^2\alpha^2-k^4\lambda^2}{k(-1+2k^3)}\\
\alpha_0=-2k^2\mu\\
\alpha_1=-2k^2\lambda\\
\alpha_2=-2k^2
\end{array}\right.\\
\mbox{ or}\\
&\left\{\begin{array}{l}
\omega=\frac{-4k^4\mu+2k^2C+k^4\lambda^2+6k^2\sigma^2\alpha^2-3\sigma^2\alpha^2}{k(-1+2k^3)}\\
\alpha_0=-\frac{2}{3}k^2\mu-\frac{1}{3}k^2\lambda^2\\
\alpha_1=-2k^2\lambda\\
\alpha_2=-2k^2
\end{array}\right.
\end{eqnarray}
 validating the relations \eqref{q1} and \eqref{q4}. $\blacksquare$
\section{Conclusion}\label{sec3}
In this work we have modified the $(G\rq{}/G)$-expansion method to
find new classes of  solutions to known important  NPDEs, in addition to   solutions obtained by the usual $(G\rq{}/G)$-expansion method. The most relevant graphical representations have been drawn to show the pertinence of the obtained analytical solutions.
\section*{Acknowledgment}
RK thanks the  University of  Abomey-Calavi and the Ministry for High Education and Scientific Research  of Benin for financial support. This work is partially supported by the ICTP through the
OEA-ICMPA-Prj-15. The ICMPA is in partnership with the Daniel
Iagolnitzer Foundation (DIF), France.

\end{document}